# Polymorphism of superionic ice

Vitali B. Prakapenka[1*], Nicholas Holtgrewe[1,2], Sergey S. Lobanov[2,3], Alexander Goncharov[2*]

[1]Center for Advanced Radiations Sources, University of Chicago, Chicago, Illinois 60637, USA.
[2]Earth and Planets Laboratory, Carnegie Institution of Washington, Washington, DC 20015, USA.
[3]GFZ German Research Center for Geosciences, Telegrafenberg, 14473 Potsdam, Germany.
*Corresponding authors: Email: agoncharov@carnegiescience.edu
prakapenka@cars.uchicago.edu

**Water is abundant in natural environments but the form it resides in planetary interiors remains uncertain. We report combined synchrotron X-ray diffraction and optical spectroscopy measurements of $H_2O$ in the laser-heated diamond anvil cell up to 150 gigapascals (GPa) and 6500 kelvin (K) that reveal first-order transitions to ices with body-centered cubic (*bcc*) and face-centered cubic (*fcc*) oxygen lattices above 900 (1300) K and 20 (29) GPa, respectively. We assigned these structures to theoretically predicted superionic phases based on the distinct density, increased optical conductivity, and greatly decreased enthalpies of fusion. Our measurements address current discrepancies between theoretical predictions and various static/dynamic experiments on the existence and location of melting curve and superionic phase(s) in the pressure-temperature phase diagram indicating a possible presence of the conducting *fcc*-superionic phase in water-rich giant planets, such as Neptune and Uranus.**

**One sentence summary (Squeezing water to superionic ice)**
*Prakapenka et al. discovered, identified, and located in the pressure-temperature diagram two superionic ices with highly mobile protons by applying synchrotron X-ray diffraction and optical spectroscopy in laser heated diamond anvil cells shedding light on the existence of these phases in planetary interiors of giant icy planets such as Neptune and Uranus.*

**Main Text:**
Ice at extreme P-T conditions experiences a dramatic modification from a hydrogen bonded molecular dipole form to extended structures (*1-6*). Upon breakdown of covalent bonding and formation of ionic solids (*e.g*. symmetric ice X (*1, 4, 7*)) the quantum and thermal proton motions become an important energy scale. This change in the energy landscape results in stability of superionic phases (*3*), which are characterized by a large proton mobility within solid oxygen sublattice and, thus, ionic conductivity. The theoretically predicted superionic ionized states of $H_2O$ are expected to appear at high pressures and interface solid ices and fluid water. The existence of superionic ices in nature has important consequences for the interior of gas giant planets, where generation of magnetic field is thought to be related to the presence of shallow fluid convective layers (*3, 8, 9*).

The phase diagram of water at high pressure is immensely controversial concerning the location of the melting line (*5, 10-18*) and the existence, structure, physical nature, and location of solid phase(s) in equilibrium with the fluid phase. The experimental and theoretical determination of the melting line vary by up to 700 K (at approximately 50 GPa) and there are no reported measurements above 90 GPa except a single point near 5000 K at 190 GPa derived from shock-wave experiments in precompressed water (*19*) (Figs. S1, S2 (*20*)). The experiments agree in that there is a sudden increase in the slope of the melting line at 20-47 GPa (*5, 10, 12, 13, 17, 18*); however, the origin of this anomaly and the location remain controversial. It has been assigned to a triple point between the fluid, ice VII, and ice X (or dynamically disordered ice VII′ (*4, 6*)) (*10, 12*), while other works suggest that it is due to the presence of superionic ice instead (*5, 21-23*) gaining the entropy compared to ice VII. Moreover, there are reports about the existence of another triple point near 20 GPa and 800 K and an additional solid phase (which melts instead of ice VII above the triple point) with unknown properties (*12*). Rigid water models and *ab initio* calculations predict the existence of plastic (with freely rotating molecules) ice phases with *bcc* and *fcc* oxygen lattices (called hereafter *bcc* and *fcc* for both plastic and superionic phases) already above 2 GPa and 300 K (*24-26*). On the other hand, above 20 GPa and 1000 K other *ab initio* simulations suggest that ice VII and fluid are interfaced by superionic phase(s) characterized by a large proton diffusivity (*3, 8, 21, 25-30*) (Fig. S2 (*20*)). The theoretically predicted superionic phases are also expected to show polymorphism above 100 GPa (*30-32*). Recent dynamic compression X-ray diffraction (XRD) experiments between 160 and 420 GPa report a transformation from a *bcc* ice X to a *fcc* superionic ice (*33*). Overall, existing experimental data and theoretical calculations show an extreme diversity concerning proton dynamics and conductivity and polymorphism of water and ices (Figs. S1, S2 (*20*)) and thus call for further experimental investigations.

Here we report the results of combined synchrotron XRD and optical spectroscopy studies in the laser heated diamond anvil cell (DAC) up to 150 GPa and 6500 K (Table S1 (*20*)) probing *in situ* structural and electronic properties of $H_2O$ ices and fluid at these conditions whereby shedding light on the phase diagram and transport properties of water at extremes. Our experiments revealed and mapped out the stability fields of two solid phases at elevated temperatures above 20 GPa that are distinct in density from the familiar ices and fluid. We assigned these phases to the theoretically predicted superionic ices based on their excessive entropy and P-T conditions of stability, which agree in the location of the melt line. The superionic nature of these phases is supported by our optical spectroscopy measurements revealing that these phases are moderately absorptive, while the same experiments detected a strong absorption threshold corresponding to the onset of electronic conductivity in fluid at about 4500 K.

Below 20 GPa, laser heating of $H_2O$ combined with an XRD probe (*20*) shows that ice VII is the only crystalline phase above room temperature and it melts at lowest reachable/detectable temperatures in good agreement with available literature data (Fig. 1). Temperature excursions at pressures above 20 GPa reveal two phase boundaries (Fig. 1). In the pressure range of ~20-60 GPa and upon heating to 900-1900 K, we detect a first-order transformation of ice VII to another *bcc* phase with lower density (Fig. 2(a), Fig. S3, Table S2 (*20*)) via an abrupt discontinuous shift of the Bragg reflections. This occurs at the P-T conditions where the majority of previous static compression measurements detected an anomaly, which was assigned to melting (*5, 14, 15, 17, 18*) (Fig. S1 (*20*)). Temperature is rising slowly with pressure along this phase transition line to 42 GPa; at this point the phase line shows an abrupt increase in slope (Fig. 1). The low-density *bcc* phase of ice discovered in our study, which we name *bcc*-SI (superionic), or ice XIX (c.f. Ref.

(*33*)) hereafter, melts along the line rising with pressure very close to that measured in Refs. (*10, 13*) up to approximately 30 GPa (Fig. S1 (*20*)).

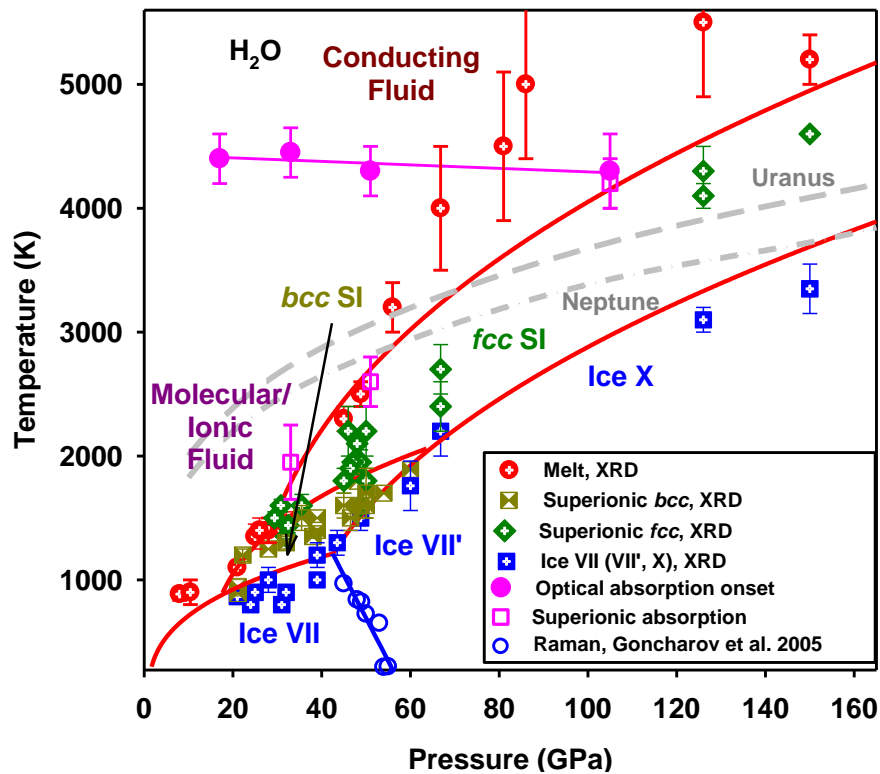

**Fig. 1**. **Phase diagram of water at extreme P-T conditions.** Solid phases are labeled after Ref. (*6*). Fluid phases are labeled following Ref. (34) and the results of this work reporting a conducting fluid. The symbols are the results of this work (20) except open blue circles, which show Raman data from Refs. (5, 6) for the phase line between ice VII and dynamically disordered ice VII′. The solid lines (guides to the eye) correspond to the proposed phase lines. The calculated isentropes of Neptune and Uranus are from Ref. (8).

At P≥29 GPa and T≥1300 K, we have observed another solid phase facing the stability field of *bcc*-SI at higher temperatures (Figs. 1, 2(b), Fig. S4-S5 (*20*)). Up to five reflections (in selected experiments) were used to identify the *fcc* structure of this phase, which we call *fcc*-SI or ice XVIII hereafter (c.f. Ref. (*33*), where only one Bragg reflection assigned to *fcc* phase was detected) (Fig. 2, Fig. S4-S5 (*20*)). At pressures above 29 GPa and below 60 GPa, the sequence of temperature induced phase transitions of $H_2O$ is the following: ice-VII(VII′) – *bcc*-SI – *fcc*-SI – fluid (Figs. 1, 2(a)). At higher pressures, the temperature stability range of *fcc*-SI increases, while that of the *bcc*-SI phase decreases and eventually vanishes above 60 GPa, where *fcc*-SI is the only stable superionic phase.

Our laser heating experiments combined with XRD detect melting via an abrupt and almost complete disappearance of the Bragg peaks (Fig. S3(b) (*20*)) and emergence of the first diffuse peak (Fig. 2(b), Fig. S3(c), S4, S6 (*20*)). However, above approximately 60 GPa only a partial melting could be observed because of large axial temperature gradients (*20*) and lack of thermal insulation. Moreover, the temperature controls deteriorated upon the transition to *bcc*-SI and *fcc*-SI phases, because of an increased laser absorption in these states; this produces a runaway increase of temperature and partial melting (Fig. 2(b), Fig. S4 (*20*)) while laser power is gradually increased. On the contrary, the temperature dropped out stepwise on tuning the laser power down.

Our measurements indicate an abrupt increase in the melting line slope above 29 GPa (Fig. 1), where the *fcc*-SI phase appears at higher temperature than *bcc*-SI and thus becomes the phase which melts (cf. Ref. (*10*)). Also, we find that the transition line between the high density *bcc* ice (VII′ or X) and *fcc*-SI and *bcc*-SI phases rises steeply above 42 GPa due to an increase in slope of the phase line related to the transition between molecular ice VII and dynamically disordered symmetric ice VII′ (*5, 6*). The latter is similar to ice X but is expected to have a bimodal proton distribution (*4*). This is qualitatively consistent with the previous observations albeit interpreted as a change in slope of the melting line (*5, 17, 18*) (Fig. S1 (*20*)) and disagrees with theoretical calculations, which predict a very flat or even negative slope of this phase line (*3, 30, 31*) (Fig. S2 (*20*)). Overall, our mapping of the P-T phase diagram signify the existence of two new ice phases *bcc*-SI and *fcc*-SI and four triple points: VII—*bcc*-SI— fluid, *bcc*-SI—*fcc*-SI— fluid, *bcc*-SI—*fcc*-SI—VII′ (X), and VII—VII′ (X)—*bcc*-SI resolving previous inconsistency in data interpretations (see Table S3 of Ref. (*20*) for the phase lines deduced here).

The unit cell volumes (densities) of newly observed *bcc*-SI and *fcc*-SI phases are quite distinct from those of ice VII (VII′ or X at higher pressure) (Fig. 3; Fig. S7, Table 4 (*20*)). The densities of *SI* phases are between those of low-temperature ices and fluid, the latter being inferred from previous experiments (*19, 22*) (Table S4 (*20*)). The temperature dependencies of density (specific volume) of SI phases could not be resolved along our experimental P-T pathways, where only moderate thermal expansion of ices VII and X reduced by the thermal pressure could be observed (Fig. 2 and Fig. S8 (*20*)). The thermal expansion effect in ices VII and X (*e.g.* measured up to 900 K in Refs. (*14, 23*)) can be sorted out because of the large and discontinuous volume expansions of *bcc* lattices upon transformation to *bcc*-SI phase (Figs. S7, S8 (*20*)). Our experiments show that the densities of *bcc*-SI and *fcc*-SI are very close to each other in the pressure range where both phases can exist (Fig. 1) and these data can be represented by the same curve (Fig. 3), indicating that these two new phases have the similar nature and evidencing that *bcc*-SI is not a thermally expended ice VII(X). The densities of SI ice (inferred from the shock velocimetry) reported along the Hugoniot (*19*) are slightly smaller compared to our data extrapolated to 185 GPa, but they agree within the error bars. On the other hand, the densities of *fcc*-SI ice inferred from the position of one XRD peak in the reverberation compression experiments (*33*) agree well with our extrapolated data. However, these data are reported at substantially lower temperatures than in our experiments (Fig. S9 (*20*)). Theoretically computed EOSs of *bcc* SI (*25, 27*) in the comparable to the experimental temperature range agree well with our results at high pressures and deviate slightly in the limit of low pressures (Fig. S7 (*20*)). The computed volume expansion due to the transformation to SI phase is comparable to our observations (*25*).

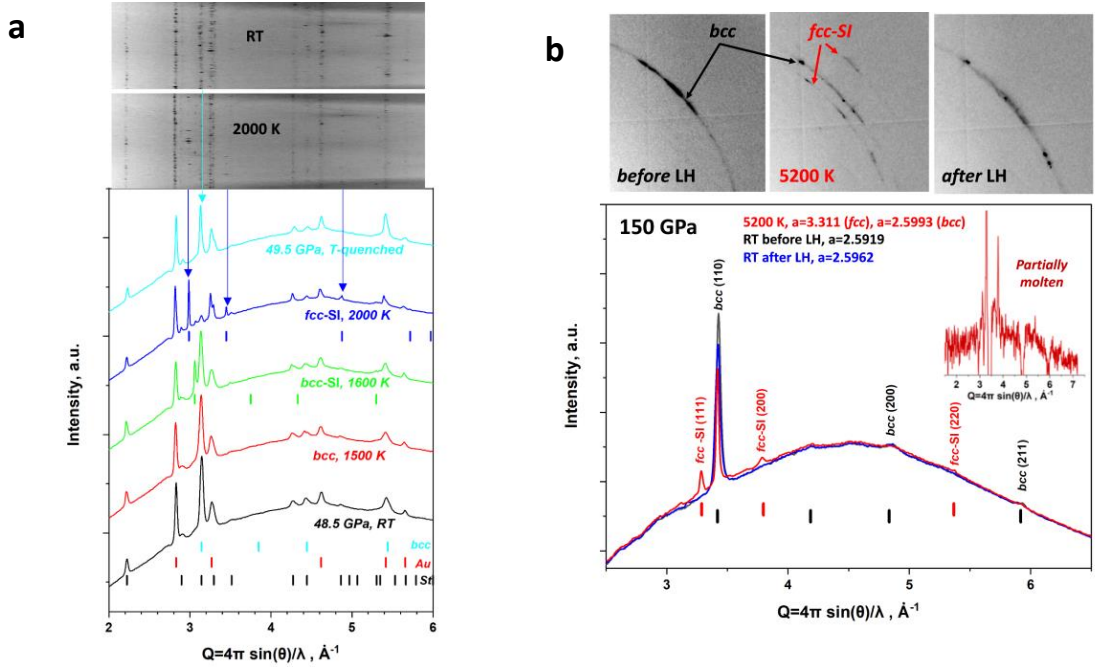

**Figure 2**. **XRD measured along near isobars at 49 GPa (a) and 150 GPa (b).** The results show the appearance of *bcc*-SI and *fcc*-SI phases at 1600 and 2000 K, respectively, at 49 GPa and only *fcc*-SI phase at 5200 K and 150 GPa. The peaks of the low-temperature phases are visible at high temperatures because of the axial temperature gradients. "*St*" stands for stishovite phase of $SiO_2$ (which was used as the thermal insulator). The transitions are fully reversible, which is seen based on XRD of the quenched to 300 K sample. The ticks correspond to the Bragg reflections of the refined structures, see Fig. S5 of Ref. (*20*) for the lattice parameters. The top inset panels are the XRD images in rectangular coordinates (cake) for 49 GPa and in polar coordinates for 150 GPa. The inset panel in (b) demonstrates diffuse scattering of partially molten water at 5200 K; it is obtained by subtracting the diffraction pattern of the quenched to 300 K sample (raw data). The X-ray wavelength is 0.3344 Å.

The phase diagram and EOSs of SI phases obtained here (*20*) can be used to understand the nature of two new ice phases, which are predicted to be superionic and appear upon heating of common dense ices above 20 GPa. Although our XRD data do not probe directly the positions of hydrogen atoms, we can get an insight on the mobility of hydrogen in SI phases via assessing the entropy change of the melting (*e.g.* Refs. (*22, 35*)). We obtain the enthalpy of fusion (latent heat), $\Delta H_f$, from the Clausius–Clapeyron relation

$$\frac{dP_m}{dT} = \frac{\Delta H_f}{T \Delta V}$$

where $P_m$ is the pressure along the melting line and $\Delta V$ is the volume change due to melting. Below 18 GPa, where molecular and dielectric ice VII melts directly to an ionized water (*22, 34*) (Fig. 1), $\Delta H_f$ increases very fast with pressure (Fig. S10 (*20*)). At higher pressures, where *bcc*-SI appears and interfaces ice VII and fluid, the melting line starts rising steeper, leading to a substantial drop of the enthalpy of fusion, which then increases with pressure gradually; a similar behavior occurs upon the appearance of *fcc*-SI phase. Fluid water is not expected to have an abrupt change in entropy over the pressure range of transitions to *bcc*-SI and *fcc*-SI, and there is no anomaly in the Δ$V$ (Fig. 3). Thus, we conclude that abrupt changes in Δ$H_f$ are due to an increase of entropy in *bcc*-SI and *fcc*-SI, especially in *bcc*-SI compared to ice VII: 62 kJ/mole vs 19 kJ/mole for transitions to *bcc*-SI and *fcc*-SI, respectively. This points to the superionic nature of *bcc*-SI and *fcc*-SI phases as predicted theoretically (*3, 8, 21, 25, 28-32, 34*) and inferred based on experimental data (*5, 19, 22, 23, 33*). The phase diagram of water is qualitatively similar to that of ammonia (*3*) in that both demonstrate the presence of superionic phases at extreme P-T conditions, however our experiments show that SI phase of water emerges at lower pressure and slightly higher temperature than in ammonia (*35*) (Fig. S11). It appears to be that the stability range of a plastic phase of water (if any) is greatly reduced compared to ammonia likely because of the presence of the strong hydrogen bonds.

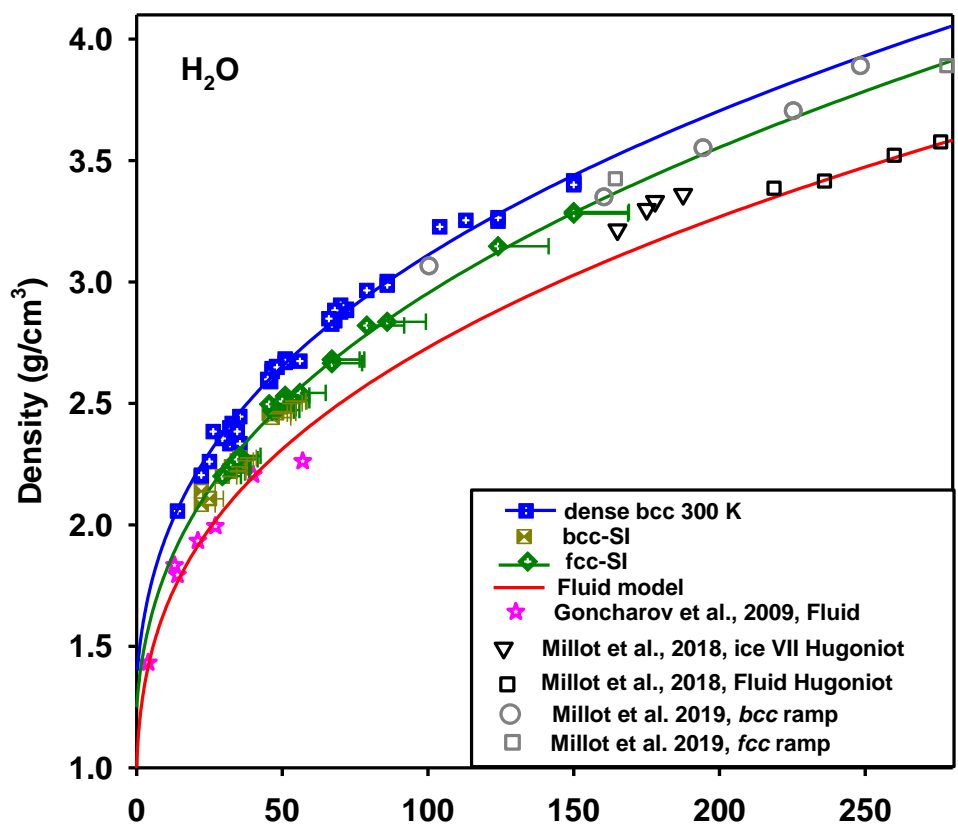

**Figure 3. Density vs P for 300 K ices, superionic phases, and fluid water**. The EOSs of combined ices VII and X at 300 K, combined *bcc*-SI and *fcc*-SI, and fluid water (all assumed temperature independent in the reported T range) are shown by solid lines; the details about the EOSs are presented in Ref. (*20*). The uncertainties of our density experiments are smaller than the symbol size. The one-directional error bars show the uncertainties in thermal pressure measurements of *bcc*-SI and *fcc*-SI (Fig. S8 of Ref. (*20*)). A detailed comparison of these data with previous experiments and theoretical calculations are shown in Fig. S7 of Ref. (*20*).

To assess the electronic properties of ices and fluid water, we directly probed the optical conductivity using visible/near IR absorption white pulsed laser (supercontinuum) spectroscopy in the pulsed laser heated DAC (*20*). The optical conductivity as low as 5 S/cm, which is near the lower limit expected for a superionic phase (*19*), could be detected in these experiments, where the thickness of the heated sample is small (a few micrometers (*36, 37*)). Our time domain absorption spectra (Fig. 4, Fig. S12 (*20*)) that were detected while the sample is pulsed heated and then naturally cooled down, show that there is a sharp temperature boundary at 4000 K, above which water is strongly absorptive with the optical conductivity >15 S/cm, similarly to our observations in $H_2$ (*37*) and $N_2$ (*38*), albeit at different temperatures. Upon cooling down the sample becomes less opaque and eventually transparent indicating a reversible transformation back to an insulating state. However, upon cooling the samples, which are heated above 4000 K at 33 and 51 GPa the sample transmission increases non-monotonously demonstrating a second transmission minimum, which we assign to optical absorption of the SI phases (Fig. S12 (*20*)). Fluid water remains non-absorptive below 4000 K as evidenced from the heating event at 17 GPa, where the sample transmission goes back to the initial level monotonously upon cooling down. The temperatures at which these absorptive states of SI ices appear are in a fairly good agreement with the phase lines determined by XRD (Fig. 1). At 17 GPa, which is close to the pressure, where the *bcc*-SI phase appears, we observed an intermittent behavior on cooling down; the sample transmission behaves regularly in some single shot events (Fig. S12 (*20*)) and shows an anomaly in another. At 105 GPa, the temperature at which the strong absorption edge is detected is very close to the melt line of *fcc*-SI, so these absorptive states of fluid and *fcc*-SI phase are difficult to distinguish. A careful examination suggests that there is a second deep transmission minimum, which we tentatively assign to *fcc*-SI phase absorption based on the time-domain radiative temperature measurements (Fig. S12 (*20*)).

The optical absorption coefficient of SI ice at 33 and 51 GPa shows an unusual increase toward the lower energy (cf. the spectra of semiconducting $H_2$ and $N_2$ (*37, 38*)), which can be attributed to superionic behavior (Fig. 4a). This behavior is likely due to highly damped low-frequency vibrational modes (*e.g.* O-H stretch), which dramatically broaden upon the transition to SI phase(s) (*e.g.* Ref. (*39*)). However, at 105 GPa, we find much stronger overall absorption and almost energy independent spectra of *fcc*-SI phase, characteristic of semiconductors with thermally activated charges (*e.g.* Ref. (*40*)). The optical conductivity of superionic and fluid phases determined here agrees well with optical experiments of Ref. (*19*) (Fig. 4b). However, we stress that the optical conductivity has a substantial ionic contribution at low temperatures (≤2500 K), where the density of thermally excited charges is small as the electronic band gap is about 4.4 eV (*40*) (c.f. DFT band gap of 2.6 eV (*41*)). In the limit of high T, our values of conductivity are in a fair agreement with previously reported shock wave electrical conductivity (*42-44*) and optical (*19, 45*) experiments. One should note, however, that unlike the presented here data, the temperatures in these shock experiments are highly uncertain (except Ref. (*19*), where it was measured radiometrically).

Our measurements clearly establish a new temperature boundary (Fig. 1) beyond which water becomes highly absorptive (likely semiconducting) similar to other materials showing a plasma transition to a conducting fluid state at similar P-T conditions (*36-38*). These results qualitatively agree with previously reported shock wave experiments (*44, 45*), however, our experiments that measure temperature directly suggest somewhat lower temperatures. Moreover, unlike shock wave experiments, our optical spectroscopy measurements in the laser heated DAC are capable of probing a wide range of P-T conditions determining the ionization and superionic phase boundaries directly. These experiments are uniquely performed at much lower temperatures than in shock experiments (*19*) revealing the absorptive nature of SI and fluid phases consistent with their ionic and electronic conductivity mechanisms. In this regard, we note that the impedance measurements of Ref. (*23*) suggested much lower temperatures for superionic states within the stability range of ice VII (VII′, X), while our experiments show an abrupt change into a superionic state along the phase line.

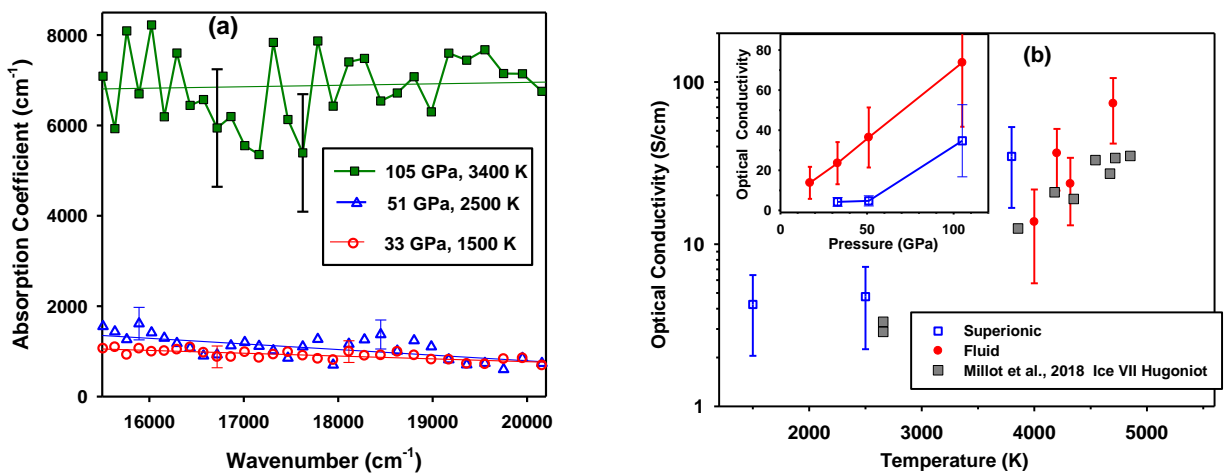

**Figure 4. Optical spectroscopy data of SI phases and fluid water**. (a) Optical absorption spectra at various P-T conditions; the error bars represent an uncertainty in the optical signal intensity. (b) Optical conductivity determined here using a broadband spectroscopy in comparison with the results at 532 nm of Ref. (*19*). The sample thickness was determined approximately using the finite element calculations (see Ref. (*20*) for more detailed information). The error bars for the conductivity values in (b) reflect this uncertainty.

Contrasting to previous static and dynamic experiments, our work provides a clear characterization of the phase and electronic states of water probed at *in situ* P-T conditions with synchrotron XRD combined with direct optical diagnostics. An experimental discovery of *bcc*-SI or ice-XIX phase reconciles previous experimental and theoretical contradictions in the position and shape of the melting line. The high-pressure *fcc*-SI (or ice-XVIII) phase has been indisputably identified here by observations of up to five Bragg reflections (Fig. S5) (cf. one reflection of Ref. (*33*)). The existence of two superionic phases is in a good agreement with the theoretical predictions (*27, 31, 32*) (Fig. S2 (*20*)), albeit our experiments find their P-T stability domains differently, likely because resolving the phase boundary between *bcc*-SI and *fcc*-SI remains a challenge for the theory (*30, 31*). Our *in situ* synchrotron XRD experiments (Figs. 2, S3-S5 (*20*)) clearly show that *fcc*-SI forms at higher T than *bcc*-SI and dominates at high P, while theories suggest that the stability of *fcc*-SI is almost solely P driven (except Ref. (*30*)). Furthermore, we show that *bcc*-SI is stable at as low as 20 GPa (in excellent agreement with theory (*3, 25, 30*) (Fig. S2 (*20*)) revealing that fluid

Hugoniot pathway only barely misses this phase (Fig. S9 (*20*)), while shock compression of even slightly precompressed water (<1 GPa as in Ref. (*45*)) should be able to probe it.

Our XRD results for SI phases extrapolated to higher pressures are consistent with that of laser shock experiments of Ref. (*19*) (Fig. S2) in the location of the melting line near 190 GPa and 5000 K and the stability domain of a SI phase (Fig. S9 (*20*)). However, the most recent reverberating shock experiments (*33*) determined much lower temperatures of stability of *fcc*-SI state (Figs. S2, S9 (*20*)), which are definitely inconsistent with our direct temperature determination. Setting aside possible temperature metrology problems (temperature was determined by model calculations in Ref. (*33*)), we propose that these nanosecond long experiments have not been able to probe the thermodynamic equilibrium states (*46*). The compression pathways of Ref. (*33*) drive the sample through the stability domain of either *bcc*-SI or *fcc*-SI depending on the strength of initial shock (Fig. S9 (*20*)). The sample was probed instantaneously by XRD at the later time (2-5 ns) where *bcc*-SI or *fcc*-SI phases (judging from their densities even though some of them are closer to ice X, Fig. S7 (*20*)) could remain metastable phase in the stability field of ice X. We emphasize that our static *in situ* experiments are crucial for understanding the phase diagram of water at extreme P-T conditions.

Our XRD and optical spectroscopy experiments established new P-T domains of superionic ices and constrained their optical conductivity values. We have also determined a temperature boundary of a large electronic conductivity of fluid water. These data allow us to address an important question about a possible contribution of water phases to the generation of the non-dipolar non-axisymmetric magnetic fields of Uranus and Neptune. Numerical dynamo simulations found that the magnetic fields of these planets are generated in a relatively thin and shallow conducting fluid shell (down to one-third of planetary radius) above a stably stratified interior (*9, 47*). We uphold this view as the P-T boundaries of $H_2O$ phases established here are consistent with fluid water in the upper third of Uranus and Neptune. At greater depths water transitions to a solid *fcc*-SI of $H_2O$ at 56(71) GPa corresponding to 74(67) % of the Uranus (Neptune) planetary radius (*48*), which may allow for the stably-stratified interior. Future studies addressing the conductivities and viscosity of superionic ices will further our understanding of the interiors of Uranus and Neptune.

In conclusion, our experiments demonstrated the existence and location of the long sought superionic phases in the phase diagram of water. We determined that there are two cubic SI phases with *bcc* and *fcc* symmetry of the oxygen sublattice (ices XIX and XVIII, respectively). Their densities are intermediate between those of low-temperature ices and fluid water. The superionicity is inferred both from the analysis of the Clausius–Clapeyron equation, which shows an excessive entropy of these phases, and from their increased optical conductivity. Our experiments reconciled previous contradictory interpretation of static and dynamic experiments as well as disagreements with theoretical calculations concerning the location of the melting line and stability fields of superionic phases. The unique state and properties of superionic phases of $H_2O$, which we confirm to be stable at super-extreme P-T conditions inside Uranus and Neptune, should be considered in modeling the interior structure and dynamics of these icy planet.

**Acknowledgments:** Porous carbon samples were received from Dr. Maria E. Fortunato and Professor Kenneth S. Suslick, University of Illinois at Urbana-Champaign. This work was performed at GeoSoilEnviroCARS (The University of Chicago, Sector 13), Advanced Photon Source (APS), Argonne National Laboratory. We thank Zachary Geballe for useful comments.
**Funding:** GeoSoilEnviroCARS is supported by the National Science Foundation – Earth Sciences (EAR – 1634415) and Department of Energy- GeoSciences (DE-FG02-94ER14466). This research used resources of the Advanced Photon Source, a U.S. Department of Energy (DOE) Office of Science User Facility operated for the DOE Office of Science by Argonne National Laboratory under Contract No. DE-AC02-06CH11357. The work at Carnegie was supported by the NSF (Grant Nos. DMR-1039807, EAR/IF-1128867, and EAR-1763287), the Army Research Office (Grant Nos. 56122-CH-H and W911NF1920172), the Deep Carbon Observatory, and the Carnegie Institution of Washington. S.S.L. acknowledges the support of the Helmholtz Young Investigators Group CLEAR (VH-NG-1325).
**Author contributions:** V.B.P. and A.F.G. conceived the experiments; V.B.P., N.H., S.S.L., and A.F.G. designed the experiments; V.B.P., N.H., and S.S.L. executed the experiments; V.B.P., N.H., and A.F.G. analyzed the data; A.F.G. and V.B.P. wrote the manuscript; and all authors reviewed and discussed the manuscript during preparation.
**Competing interests:** All authors declare no competing interests.
**Data and materials availability:** All data are available in the manuscript or the supplementary information.

**SUPPLEMENTARY INFORMATION**
Materials and Methods
Fig. S1-S15
Tables S1-S4
References (1-41)